\documentclass[fleqn,usenatbib,useAMS,twocolumn]{aastex63}

\usepackage{ulem}
\usepackage{amsmath}
\usepackage{tensor}     
\usepackage{graphicx}   
\usepackage{bm}         
\usepackage{xcolor}     
\usepackage{color}      
\usepackage[section]{placeins} 
\usepackage[utf8]{inputenc} 

\newcommand{\nc}{\newcommand*} 
\nc{\figurewidth}{3.2in}
\nc{\xbar}{\bar{x}}
\nc{\rhoeq}{\rho_{\mathrm{eq}}}
\nc{\zeq}{z_{\mathrm{eq}}}
\nc{\tla}{\tilde{\lambda}}
\nc{\dt}{\delta}
\nc{\Dt}{\Delta}
\nc{\vj}{\vec{j}}
\nc{\vl}{\vec{l}}
\nc{\hx}{\hat{x}}
\nc{\hy}{\hat{y}}
\nc{\bj}{\bm{j}}
\nc{\mJ}{\mathcal{J}}
\nc{\mP}{\mathcal{P}}
\nc{\Msun}{M_\odot}
\nc{\app}{\approx}
\nc{\av}[1]{\langle #1 \rangle}
\nc{\eq}[1]{Eq.~\eqref{#1}}
\nc{\al}{\alpha}
\nc{\Xstar}{X_{\ast}}
\nc{\seq}{\sigma_{\mathrm{eq}}}
\nc{\fpbh}{f_{\mathrm{pbh}}}
\nc{\vth}{\vec{\theta}}
\nc{\vla}{\vec{\lambda}}
\nc{\vd}{\vec{d}}
\nc{\Mmin}{M_{\mathrm{min}}}
\nc{\rmd}{\mathrm{d}}
\nc{\mmin}{{m_{\mathrm{min}}}}
\nc{\mmax}{{m_{\mathrm{max}}}}
\nc{\mR}{\mathcal{R}}
\nc{\tmR}{\tilde{\mathcal{R}}}
\nc{\s}{\sigma}
\nc{\ogw}{\Omega_{\mathrm{GW}}}
\nc{\addref}{[\textcolor{red}{add ref}] }
\nc{\Om}{\Omega}
\nc{\gpcyr}{\mathrm{Gpc}^{-3}\,\mathrm{yr}^{-1}}
\nc{\Eq}[1]{Eq.~\eqref{#1}}
\nc{\Fig}[1]{Fig.~\ref{#1}}
\nc{\Table}[1]{Table~\ref{#1}}
\nc{\lvc}{LIGO/Virgo} 
\nc{\Sec}[1]{Sec.~\ref{#1}}
\nc{\eg}{\textit{e.g.~}}
\nc{\SNR}{\mathrm{SNR}}

\def\p{\partial}

\def\({\left(}
\def\){\right)}
\def\[{\left[}
\def\]{\right]}

\def\e{\begin{equation}}
\def\q{\end{equation}}
\def\m{\begin{eqnarray}}
\def\n{\end{eqnarray}}

\begin{document}

\title{The dynamics and gravitational-wave signal of a binary flying closely by a Kerr supermassive black hole}

\author[0000-0002-1459-0879]{Zhongfu Zhang}
\affiliation{Department of Astronomy, School of Physics, Peking University, 100871 Beijing, China}
\author[0000-0003-3950-9317]{Xian Chen}
\affiliation{Department of Astronomy, School of Physics, Peking University, 100871 Beijing, China}
\affiliation{Kavli Institute for Astronomy and Astrophysics at Peking University, 100871 Beijing, China}
\correspondingauthor{Xian Chen}
\email{xian.chen@pku.edu.cn}

\begin{abstract}
Recent astrophysical models predict that 
stellar-mass binary black holes (BBHs) could form and coalesce 
within a few gravitational radii of a supermassive  black hole (SMBH).
Detecting the gravitational waves (GWs) from such systems requires numerical tools
which can track the dynamics of the binaries while capturing 
all the essential relativistic effects. This work develops upon our earlier
study of a BBH moving along a circular orbit in the equatorial plane of a Kerr
SMBH. Here we modify the numerical method to simulate a BBH falling toward the SMBH
along a parabolic orbit of arbitrary inclination with respect to the equator.
By tracking the evolution in a frame freely falling alongside the binary, we find that the eccentricity
of the BBH is more easily excited than it is in the
previous equatorial case, and that the cause is the asymmetry of the tidal tensor
imposed on the binary when the binary moves out of the equatorial plane. 
Since the eccentricity reaches maximum around the same time that the BBH
becomes the closest to the SMBH, multi-band GW bursts could be produced which are simultaneously
detectable by the space- and ground-based detectors.
We show that the effective spins of such GW events also undergo significant variation due to the
	rapid reorientation of the inner BBHs during their interaction with
SMBHs. These results demonstrate the richness of
three-body dynamics in the region of strong gravity, and highlight the necessity of building new
numerical tools to simulate such systems. 
\end{abstract}


\section{Introduction} \label{sec:intro}

The interaction between supermassive black holes (SMBHs) and binary compact
objects, such as stellar-mass binary black holes (BBHs) or double neutron stars (DNSs), 
can produce various types of gravitational-wave (GW) sources.  (i) If a
BBH penetrates deeply into the potential well of a SMBH, it is likely tidally
disrupted, and the corresponding distance from the SMBH is called the ``tidal
radius'' \citep{Hills88}. In many cases, the tidal disruption will leave one of
the stellar-mass black holes (BHs) tightly bound to the SMBH, forming a GW source known as
the extreme-mass-ratio inspiral \citep{MillerEtAl05}, which is one of the
major targets of the Laser Interferometer Space Antenna (LISA, \citealp{AstrophLISA2023}).  (ii) Later numerical simulations showed that not all BBHs
reaching the tidal radius of a SMBH are disrupted. A substantial fraction can
survive, but the eccentricities of the surviving binaries can be excited to
large values during the tidal interaction
\citep{addison_gracia-linares_2019,Joseph2019MNRAS}. The high eccentricity
enhances GW radiation, causing the binaries to coalesce later and become
potential targets of the Laser Interferometer Gravitational-wave Observatory
(LIGO) and the Virgo detectors \citep{GWTC1,GWTC2,GWTC3PhRvX2023}.
(iii) If the distance is not as close as the tidal radius (and not
significantly larger either), the interaction could result in a capture of the
binary by the SMBH \citep{Hills91}. 
The captured binary could also be driven to a high eccentricity 
by the nearby SMBH and coalesce.
If the binary is a BBH, when it coalesces the triple system could 
emit GWs which are simultaneously
detectable by LISA as well as LIGO/Virgo
\citep{2018CmPhy...1...53C,han_chen_2018}.  (iv) At even larger distances, the
interaction between an SMBH and a BBH becomes more stable. Such long-term
interaction could periodically excite the BBH to high eccentricities
via a Von Zeipel-Lidov-Kozai mechanism
\citep{antonini10binary-SMBH,antonini_perets_2012}.  Consequently, the
lifetimes of the BBHs are reduced
\citep{petrovich_antonini_2017,meiron_kocsis_2017,hoang18,hamers18,fragione19}.
This mechanism could have contributed a large fraction of the BBHs
detected by LIGO/Virgo \citep{takatsy19,wang19,zhang19NSC,ArcaSedda20}.

More recent studies have revealed cases in which the BBHs can reach a distance
smaller than ten Schwarzschild radii ($R_S$) from a SMBH. First, the
aforementioned mechanism of tidal capture normally deposit BBHs on highly
eccentric orbits around a SMBH. The pericenter distances of the orbits can be
comparable or smaller than $10R_S$ when the population of the most compact BBHs
are considered \citep{han_chen_2018,addison_gracia-linares_2019}. Second, it is
well known that the accretion disk of an active galactic nucleus (AGN) is a
breeding ground of BBHs \citep{baruteau11,mckernan_ford_2012} and DNSs
\citep{cheng99}.  While the majority of the BBHs form and merge relatively far
away from the SMBH
\citep{2016ApJ...819L..17B,bartos_kocsis_2017,stone_metzger_2017,secunda19,li22,li23,DeLaurentiis23},
a small fraction of them migrate fast in the accretion disk and can reach the
inner edge of the disk \citep{chen_li_2019,tagawa20,peng21,peng23}.

At such a small distance from the SMBH, the dynamical evolution of the BBH is
strongly affected by relativistic effects. For example, the apsidal precession
of the ``outer orbit'' (the orbital motion of the BBH around the SMBH) becomes
important. If this precession resonates with  the apsidal precession of the
inner BBH, the eccentricity of the inner binary could be excited \citep{liu20}.
Moreover, the BBH starts to feel the frame-dragging effect if the central SMBH
is spinning.  Such an effect could enhance the Von Zeipel-Lidov-Kozai
oscillation of the BBH \citep{liu19spin,2019ApJ...887..210F,liu22}.  The
emergence of these effects also motivated a recent development of an effective
field theory to tackle the relativistic three-body problem in general
\citep{kuntz21,kuntz23}.

Besides dynamics, the GW signal of a merging BBH (or DNS) is also affected by
the presence of a nearby SMBH. First, the Doppler and gravitational redshifts
induced by the SMBH, which normally are not considered in the data analysis,
can make the merging binary appear more massive and more distant in the
detector frame \citep{chen_li_2019,vijaykumar22,Zhangxy2023MNRAS}. Second, the
high velocity of the binary around the SMBH beams the GW radiation, which can
affect the measurement of the distance
\citep{torres-orjuela_chen_2019,ale23,yan23}.  Third, because of its motion
around the SMBH, the center-of-mass (c.m.) velocity of the binary varies with
time (called ``peculiar acceleration'').  The variation modulates the redshift
\citep{bonvin17,meiron_kocsis_2017,inayoshi_tamanini_2017,tamanini20} as well
as the effective viewing angle \citep{Torres-OrjuelaEtAl2020} of the binary,
resulting in a phase shift which is potentially detectable by LISA. If the
binary is located at the inner edge of the accretion disk of an AGN, the phase
shift accumulated during the final few seconds of the merger can be large
enough to be detectable by ground-based detectors \citep{vijaykumar23waltzing}.
Fourth, the curved spacetime of a SMBH can bend the trajectories of the GWs
emitted from the merging binary
\citep{campbell73,lawrence73,ohanian73,kocsis13,dorazio20,yu21,gondan22,oancea22}.
Finally, the GWs could also be amplified by a Penrose-like process
\citep{gong21} or by resonating with the qusi-normal modes of the central SMBH
\citep{cardoso21}.

Despite the increasing interest in studying the dynamics and GW signal of a
binary in the vicinity of a SMBH, one problem remains and becomes more
prominent. The conventional tools of simulating three-body dynamics becomes
insufficient as the triple system becomes more relativistic.  On one hand, the
commonly used post-Newtonian (PN) formalism \citep{will14,2017PhRvD..96b3017W}
breaks down because the presumption, that the velocities of the bodies are much
smaller than the speed of light, is invalid in the current case ($r\la 10R_S$).  On the
other hand, the outer orbit is typically $10^4$ times bigger than the size of
the inner binary, if not greater. Such a large dynamical range makes it
intractable to solve the problem by full numerical relativity
\citep[e.g.][]{bai11}.

Several recent works made a first step towards solving the above problem.  The
key idea is that in a frame freely falling alongside the inner binary, by the
equivalence principle, the dynamics is much simpler.  In this free-fall frame (FFF), the
equations of motion of the binary are determined mainly by its self-gravity,
plus a weak perturbation induced by the spacetime curvature of the SMBH
\citep{Gorbatsievich10,komarov18}. The perturbation behaves like an
electromagnetic force, known as the gravito-electromagnetic (GEM) force
\citep{mashhoon03}. Using the GEM formalism and assuming a qusi-circular outer
orbit, \citet{paper2022} showed that a BBH very close (several gravitational
radii) to a Kerr SMBH cannot move along a geodesic line. Moreover, the 
inner orbital eccentricity of the BBH
could evolve to a high value even though the binary is coplanar with the outer
orbit.  \citet{Camilloni2023PhRvDGEM} assumed a circular geodesic motion for
the BBH and used a double-averaging technique to derive the long-term evolution
of the binary in the comoving frame.  Most recently,
\citet{Maeda2023PhRvDGEM,maeda23} used a Fermi-Walker transport to correct for
the deviation of the c.m. of the binary from geodesic motion, and they
reanalyzed the criterion for the stability of a BBH on a quasi-circular orbit
in the equatorial plane of a Kerr SMBH.

In these previous works, the outer orbits are
usually chosen to be coplanar with the equatorial plane of the Kerr SMBH and
nearly circular, so that the curvature tensor in the FFF of the BBH takes
simple form.  However, we have mentioned above that the outer orbits can be
highly elongated, for example, if BBHs are tidally captured by SMBHs
\citep{2018CmPhy...1...53C,addison_gracia-linares_2019}.  Such orbits can send
BBHs to even closer distances from the SMBH, not limited by the innermost
stable circular orbit. As the distance decreases, the curvature tensor
increases and varies faster with time.  Elongated outer orbits are not
restricted in the equatorial plane either. Outside the equatorial plane, the
curvature tensor becomes more asymmetric \citep{bardeen72}. This paper aims at
studying these new effects on the dynamical evolution of the BBHs.

The paper is organized as follows.  In Section~\ref{sec:meth} we describe the
theoretical framework of simulating the evolution of a BBH in its FFF. Based on
the observation that the perturbation by the Kerr background induces GEM forces
in the FFF, we argue that the asymmetric and non-diagonal forces can drive the
dynamical evolution of the BBH.  In Section~\ref{sec:result} we carry out
numerical simulations to verify our analytical argument. We pay special attention
to the properties of the surviving binaries, including their orbital elements
and lifetimes after the interaction.  In Section~\ref{sec:dis} we discuss the
possible observational signatures imprinted in the GW signal of such a triple
system, as well as the caveats for future improvement.  Throughout the paper,
we use geometrized units where $G = c =1$.

\section{Numerical method} \label{sec:meth}

\subsection{Brief review of the method}

The method used here is adopted from our earlier work \citep{paper2022}.  For
the completeness of this work, we briefly review the major steps.  The
system of our interest has a clear hierarchy. On one hand, the SMBH has a
typical mass of $10^6\lesssim M/M_\odot\lesssim10^9$. The spacetime
close to the SMBH has a typical curvature radius of $\rho\sim M$. On the other
hand, the BBH, which we refer to as the ``inner binary'', has a semi-major axis
of $a\simeq(10^3-10^4)m_{12}$
\citep{2018CmPhy...1...53C,addison_gracia-linares_2019,peng21}, where
$m_{12}=m_1+m_2\simeq{\cal O}(10)M_\odot$ is the total mass of the binary. We
find that $a/\rho\ll1$.  This ratio indicates that spacetime is sufficiently
flat within the vicinity of the BBH. In this case, the dynamics is sufficiently
simple in a frame falling freely together with the BBH. In this frame, the
binary evolves mainly according to its self-gravity, except for a weak
perturbation induced by the small background curvature. Taking advantage of the
above hierarchy, we divide our calculation into two parts. 

First, we compute the geodesic motion of a free-fall observer in the Kerr
metric. The computation is carried out in the Boyer-Lindquist coordinates
$(t,r,\theta,\varphi)$. Following the convention of three-body dynamics, we
call this geodesic the ``outer orbit''. To facilitate the following
calculation,  we construct a local frame centered on the free-fall observer
such that sufficiently close to the observer the metric is approximately
Minkowskian.  The corresponding coordinates are known as the Fermi normal
coordinates \citep{Fermi1922,FNC63} and we denote them as $(\tau,
\mathbf{x})=(\tau, x, y, z)$, where $\tau$ is the proper time inside this
FFF. 

Second, we place a BBH of our interest in the FFF. Initially,  
the c.m. of the BBH coincides with the origin of the FFF, 
and they share the same velocity in the Boyer-Lindquist coordinates.
The subsequent evolution of the BBH in the FFF is determined by 
\begin{equation}
	m_a\frac{d^2\mathbf{x}_a}{d\tau^2}=m_a m_b\frac{\mathbf{x}_b-\mathbf{x}_a}{|\mathbf{x}_a-\mathbf{x}_b|^3}+\mathbf{F}_a(\tau,\mathbf{x}_a,\mathbf{v}_a)+\mathbf{F}_{\rm PN},
\label{eq:eom}
\end{equation}
where $a, b=1,2$ denote the two stellar-mass BHs.  Here, $\mathbf{F}_a$ is the
GEM force induced by the weak background curvature \citep{mashhoon03}.  The
last term $\mathbf{F}_{\rm PN}$ accounts for PN corrections, which becomes
important when the two stellar BHs are close to each other.  We have included the PN terms up
to the 2.5 order \citep{Blanchet2014} so that we can simulate the merger of the
BBH due to GW radiation. 

The GEM force in Equation~(\ref{eq:eom}) can be written as
\begin{equation}
	\mathbf{F}=-m\mathbf{E}-2m\mathbf{v}\times \mathbf{B} \label{eq:F}
\end{equation}
\citep{mashhoon03}, where $m$ is the rest mass of an object and
$\mathbf{v}:=d\mathbf{x}/d\tau$ is its velocity relative to the FFF. 
The gravito-electric (GE)  and gravito-magnetic (GM) fields, 
$\mathbf{E}$ and $\mathbf{B}$, are calculated with
\begin{eqnarray}
	E_i(\tau,\mathbf{x})&=&R_{0i0j}(\tau)x^j,\label{eq:E}\\
	B_i(\tau,\mathbf{x})&=&-\frac{1}{2}\epsilon_{ijk}{R^{jk}}_{0l}(\tau)x^l,\label{eq:B}
\end{eqnarray}
where $R_{0i0j}$ and ${R^{jk}}_{0l}$ are the components of the Riemann tensor
in the FFF, $\epsilon_{ijk}$ is the Levi-Civita symbol, and $i$, $j$, $k$, and
$l$ are spatial indices which take the values $1, 2, 3$. In the following, we
refer to the aforementioned two Riemann-tensor components as $R_{ij}$ and
${R^{jk}}_{l}$ for simplicity.
 
\subsection{Implementing a parabolic outer orbit}

The previous works have focused on the BBHs on circular orbits around Kerr SMBHs
\citep{paper2022,Maeda2023PhRvDGEM,Camilloni2023PhRvDGEM}. Describing such an orbit
is relatively simple, because it is restricted in the equatorial plane of the
SMBH and keeps a constant distance from the central SMBH. A parabolic outer
orbit is more complicated.  It allows three constants of motion, the specific
energy $E=1$, the component of angular momentum along the spin axis of the SMBH
$L_z$, and the Carter constant $Q$ \citep{PhysRev.174.1559}.  The latter two
need to be specified in this work.

In practice, we substitute $Q$ using the pericenter distance $r_p$ of the orbit.
The relationship between $r_p$ and the constants of motion is   
\begin{equation}
r_p^3   
+\left[(sM  - L_z)^2 + Q\right] r_p 
- \frac{L^2 r_p^2}{2M}
	- \frac{s^2M Q}{2}=0,\label{eq:rp}
\end{equation}
where $s\in[0,1)$ is the spin parameter of the SMBH and $L^2=Q+L_z^2$.  
Given $r_p$, the choice of $L_z$ is not totally free. The condition
$Q\ge0$ requires that
\begin{eqnarray}
    \frac{-2s-\sqrt{2r_p\Delta}}{(r_p-2)}\le L_z\le
    \frac{-2s+\sqrt{2r_p\Delta}}{(r_p-2)},
    \label{eq:Lz}
\end{eqnarray}
where $\Delta=r_p^2-2Mr_p+M^2s^2$. 

Having chosen the constants of motion, we also need to specify two Boyer-Lindquist coordinates,
$r$ and $\theta$, to fix the starting point of the parabolic orbit.
For $r$, it cannot be smaller than the radius of the 
marginally bound circular orbit $r_{\rm mb}$ \citep{bardeen72}.
For $\theta$, it is restricted in the region where  $(\cos\theta)^2\le Q/(Q+L_z^2)$,
so that the solution to the coordinate velocity $d\theta/dt$ exits.

\begin{figure}[ht!]
\includegraphics[width=0.5\textwidth]{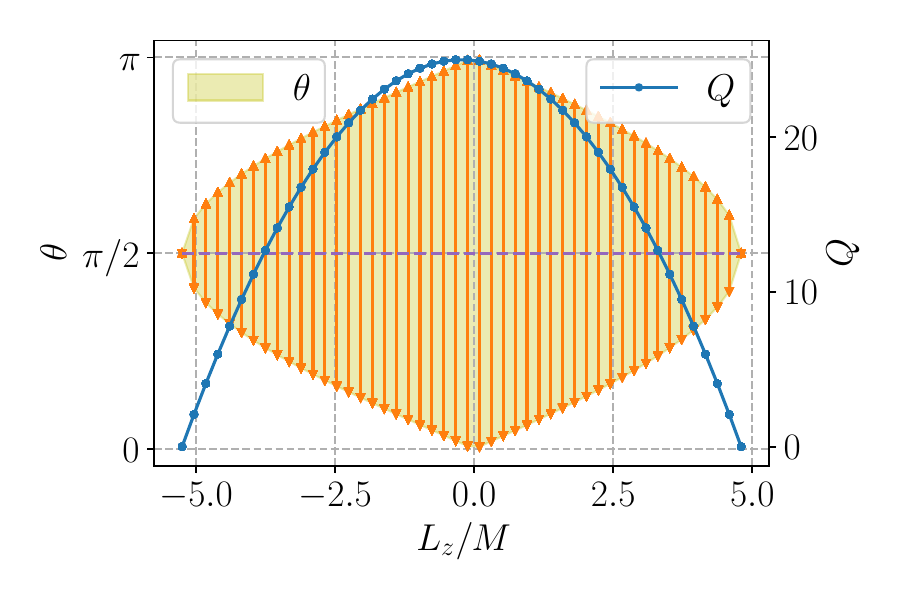}
\caption{The range of the inclination angle $\theta$ (shaded area with arrows) and the value of the Carter constant $Q$ (blue curve with dots) 
as a function of $L_z$ , the angular-momentum component projected on the spin axis of the
central Kerr SMBH. We have assumed $r_p=10M$ and $s=0.9$ in the plot.
\label{fig:rangeTheta}}
\end{figure}

To illustrate the complexity of a parabolic orbit,
Figure~\ref{fig:rangeTheta} shows the allowed range of $\theta$,
as well as the value of $Q$, as a function of $L_z$.  In the plot we have set
$r_p=10M$ and $s=0.9$.  The curve of $Q$ is axisymmetric, but the axis of
symmetry is offset from $L_z=0$ due to the spinning of the SMBH.  It is also
clear that the geodesic can leave the equatorial plane ($\theta=\pi/2$)
when $Q\neq0$.

\subsection{Tidal forces in the free-fall frame}

Having set up a parabolic orbit, we now calculate the tidal forces in the frame
freely falling along the parabola.  In this work we will focus on the GE
forces. We can neglect the GM ones for the following two reasons. (i) The GM
forces, according to Equation~(\ref{eq:F}), are smaller than the GE ones by a
factor of $v/c$. In our problem where $a\simeq(10^3-10^4)m_{12}$, we have
$v/c\sim{\cal O}(10^{-2})$. (ii) In our earlier work we found that the
dynamical effect of GM forces is accumulative \citep{paper2022}. In order for
such an effect to build up, the inner binary has to complete tens of
revolutions. However, the parabolic encounter considered here is much quicker.
Typically, the inner binary has time to complete only a couple of revolution during
the pericenter passage.
 
Since GE forces are commonly known as the ``tidal forces'', the corresponding
Riemann tensor, i.e., $R_{ij}$ in Equation~(\ref{eq:E}), are also called the
tidal tensor.  The tidal tensor in an FFF moving along an arbitrary time-like
geodesic has been derived in \citet{Marck1983}.  Its components are the
simplest when written in a local inertia frame (LIF), which differs from the
FFF by a rotation \citep[see our earlier work,][for explanation]{paper2022}. In
the LIF, the $R_{13}$ and $R_{23}$ components and their symmetric parts (note
that $R_{ij}$ is symmetric) will vanish. 

\begin{figure}[ht!]
\includegraphics[width=0.5\textwidth]{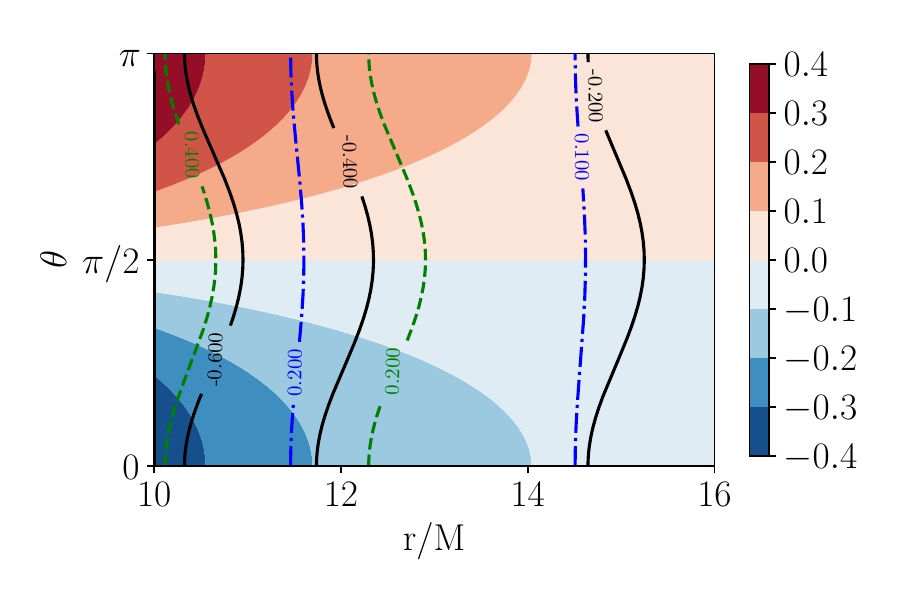}
\caption{Tidal forces in the local inertia frame as a function of the
Boyer-Lindquist coordinates $r$ and $\theta$. 
The color map shows the force induced by the off-diagonal component 
$R_{12}$ of the tidal tensor. In comparison, the black solid, green dashed, and blue dot-dashed
contours correspond to the forces induced by, respectively, $R_{11}$, 
$R_{22}$, $R_{33}$. 
The value shown here is in unit of the self-gravity of the BBH, i.e., $-m_{12}/a^2$.
In the plot, we have assumed 
$M=4\times10^6M_\odot$, $m_1=15M_\odot$, $m_2=10M_\odot$, $a=2\times10^4m_{12}$,
$E=1$, $r_p=10M$, $s=0.9$, and $L_z=2.2M$. 
\label{fig:compF12}}
\end{figure}

While the tidal tensor of a circular outer orbit (the focus of our previous
work) contains only diagonal components in the LIF and remains constant, the
tidal tensor of a parabolic geodesic differs drastically.  First, the
off-diagonal component $R_{12}$ in the LIF will appear as soon as the geodesic
leaves the equatorial plane of the SMBH. This property is illustrated in
Figure~\ref{fig:compF12}, where the color map shows the magnitude of the tidal
force, $R_{12}a$, divided by the self-gravity of the inner binary,
$-m_{12}/a^2$. We can see that the $R_{12}$ component is non-zero outside the
equatorial plane, and it can produce a tidal force as large as $40\%$ of the
self-gravity of the inner BBH when $r$ approaches the pericenter.  This tidal
force is comparable to the ones resulting from the diagonal components of the
tidal tensor (see the contours).  Such a large off-diagonal component will
impose an additional tidal torque which is absent from the earlier studies of
the BBHs in the equatorial plane. 

\begin{figure}[ht!]
\includegraphics[width=0.5\textwidth]{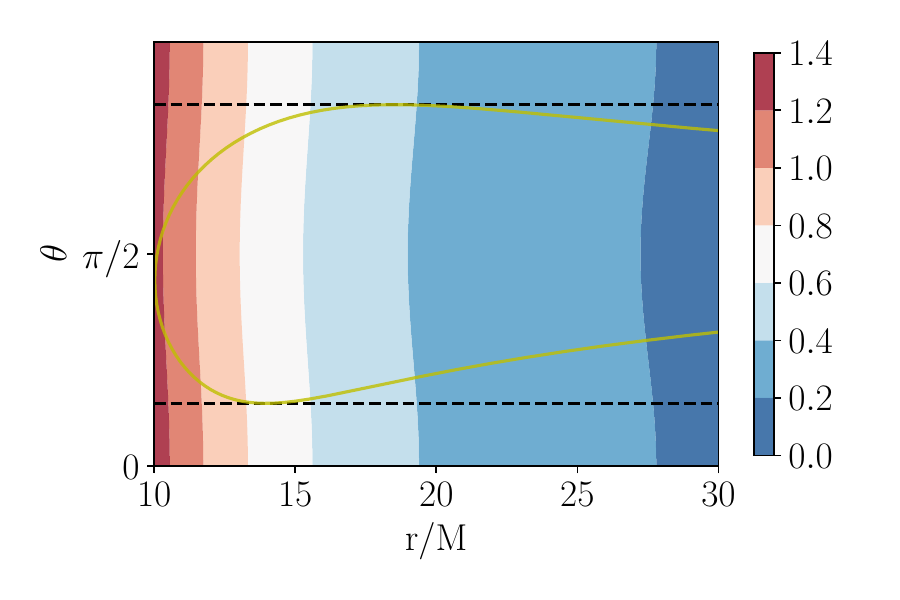}
\caption{Rotational velocity of the tidal tensor in the FFF as 
a function of the $r$ and $\theta$ coordinates. The value is in unit of $\sqrt{M/r_p^3}$.
The yellow curve shows the projection of a parabolic orbit, which starts in the
equatorial plane at a radius of $r=100M$. The black dashed lines are the expected turning
points in the $\theta$ direction. The other parameters are the same as in Figure~\ref{fig:compF12}.
\label{fig:compPhi}}
\end{figure}

The second dynamical effect induced by a parabolic orbit is a variation of the
rotational velocity of the FFF relative to the LIF. This rotational velocity can be calculated with
\begin{equation}
\begin{aligned}
\omega=&\frac{K^{\frac{1}{2}}}{r^2+s^2M^2\cos^2\theta}\\
	\times&\left( \frac{r^2+s^2M^2-s M L_z}{r^2+K}+ \frac{s M L_z-s^2M^2  \sin ^2 \theta}{K-s^2M^2 \cos ^2 \theta}\right)\label{eq:omega}
\end{aligned}
\end{equation}
\citep[see Eq.~(46) in][]{Marck1983}, where $K=(s ME -L_z)^2+Q$ is another
definition of the Carter constant and we have used $E=1$ in the
calculation.  Figure~\ref{fig:compPhi} shows the value of $\omega$ for
different $r$ and $\theta$.  We find that the rotational velocity is determined
mainly by the $r$ coordinate, and it can vary by orders of magnitude along a
parabolic orbit. Since the axes of the LIF determine the
orientation of the tidal tensor, a variation of the rotational velocity of the
LIF will make the tidal forces in the FFF behave more irregularly.

\section{Scattering experiments and results}\label{sec:result}

\subsection{Initial conditions}\label{sec:ini}

In our fiducial model, we choose the following parameters,
$M=4\times10^6M_\odot$, $s=0.9$, $m_1=15M_\odot$, and $m_2=10M_\odot$.  
The mass of the SMBH resembles that of the Milky Way, although it
is unlikely that the SMBH in the Galactic Center is currently
accompanied by a BBH \citep{2018CmPhy...1...53C,peng21}.

For the outer orbit, we assume that initially $r_p=10M$ and $E=1$. For $L_z$,
we assume a uniform distribution within the range given by
Equation~(\ref{eq:Lz}).  Such a distribution corresponds to a random
distribution of orbital inclination with respect to the equatorial plane of the
Kerr SMBH. Moreover, initially we place the c.m. of the BBH in the equatorial
plane. It will later leave the equatorial plane if $Q>0$.
Note that the above initial conditions cannot determine
the initial radius of the outer orbit. For this reason, we choose $r_{\rm
ini}=100M$. Such a radius is far enough from the central SMBH, so that the
tidal perturbation  has not significantly affected the dynamics of the BBH.
Meanwhile, $r_{\rm ini}$ is relatively small, so that we can complete the
simulation within a reasonable amount of computational time.

For the inner BBH, we choose an initial semimajor axis of $a_0=20,000m_{12}$,
which corresponds to an orbital period of
$\tau_0=2\pi\sqrt{a_0^3/m_{12}}\simeq2,217~{\rm s}$. Since $a_0$ is much
smaller than the curvature radius of the Kerr SMBH, the precision in the
calculation of the GEM forces can be guaranteed.  For simplicity, we further
assume that the inner orbit is initially circular. The corresponding GW
radiation timescale is $5a_0^4/(256m_1m_2m_{12})\simeq 52,000$ yrs
\citep{peters_1964}, much longer than the mission duration of LISA.  As for the
stability of the BBH, we follow the convention and define a penetration factor
of $\beta=r_t/r_p$, where $r_t=(M/m_{12})^{1/3}a_{0}$ is the tidal disruption
radius derived in the Newtonian limit.  In our fiducial model, we have
$\beta\simeq0.6786$, indicating that the BBH is marginally stable. 
The inclination of the inner BBH is defined by the angle
$\iota$ between the inner orbital angular momentum of the BBH with respect to the rotating
axis of the LIF.
 
\subsection{Parameter space of the surviving BBHs}\label{sec:semimajor axis}

\begin{figure}[ht!]
\includegraphics[width=0.5\textwidth]{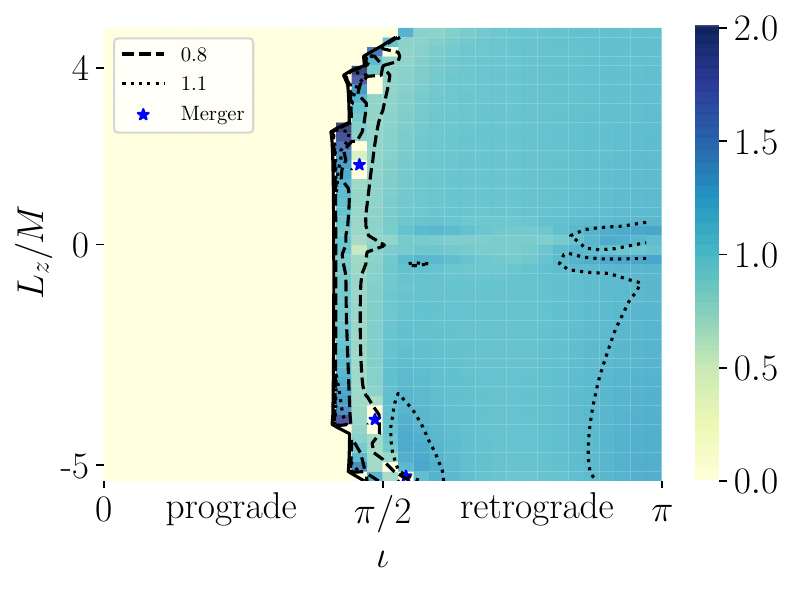}
\caption{Dependence of the final semimajor axis of the inner BBH 
on the initial values of $L_z$ and $\iota$. 
The other parameters are the same as in Figure~\ref{fig:compF12}.
The solid curve marks the boundary between
tidal disruption and survival. The dashed and dotted contours show the regions where the
final semimajor axis is, respectively, $0.8a_0$ and $1.1a_0$. The blue stars show the 
cases in which the BBHs coalesce during the pericenter passage.
\label{fig:afin}}
\end{figure}

Since the penetration factor in our fiducial model is close to unity, a large
fraction of the BBHs are tidally disrupted during their encounters with the
central Kerr SMBH.  Interesting, according to Figure~\ref{fig:afin}, the
surviving binaries occupy a specific region in the parameter space of $L_z$
and $\iota$.

The survivors reside mostly at $\iota\ga\pi/2$. The reason for their survival
is that they are counterrotating with respect to the LIF.  For
example, the rotational velocity of the BBH itself is about
$\omega_b=\sqrt{m_{12}/a^3}\simeq 5.2\times10^{-3} \rm{rad/s}$, and the
rotational velocity of the LIF relative to the FFF is
$\omega\sim\sqrt{M/r_p^3}\simeq 3\times10^{-3} \rm{rad/s}$ when the BBH
approaches the pericenter of the outer orbit.  Therefore, in the rest frame of the binary, the
tidal field could rotate at an angular velocity as high as
$\omega_b+\omega\simeq8\times10^{-3} \rm{rad/s}$.  Such a rapid variation
effectively washes out the asymmetry of the tidal field, so that tidal
disruption becomes less likely.  In the following, we refer to these BBHs with
$\iota>\pi/2$ as ``retrograde'' binaries.  On the contrary, the BBHs with
$\iota<\pi/2$ are rotating in the same sense as the rotation of the LIF, and we
call them ``prograde'' binaries. In the rest frame of a prograde BBH, the LIF
has an angular velocity of $\omega-\omega_b\simeq-2\times10^{-3} \rm{rad/s}$.
Such a slower rotation induces a more persistent tidal force on the BBH, so that
the binary is disrupted more easily. 

Figure~\ref{fig:afin} also shows that $L_z$ plays a minor role in determining
the stability of the BBHs. The weak dependence reflects the fact that at
$r=r_p$, the angular velocity of the LIF, $\omega$, is insensitive to the value
of $L_z$. Nevertheless, close to $L_z\simeq4M$ and $L_z\simeq-5M$, more BBHs
are tidally disrupted than in the other cases of $L_z$. 
The reason is that the orbits here are closer to the
equatorial plane when $|L_z|$ is greater. According to
Figure~\ref{fig:compF12}, the tidal forces induced by the diagonal components
of the  tidal tensor are larger in the equatorial plane than in the polar
regions.

We notice that \citet{Heggie1996} studied the interaction of a binary with a third body
and derived a criterion for the survival of the binary,
\begin{align}
    \mathcal{F}^{-1}(\iota)>6 \sqrt{\pi} 2^{1 / 4} \beta^{-3 / 4} \exp \left[-\frac{2 \sqrt{2}}{3} \beta^{-3 / 2}\right],
    \label{eq:cosiN}
\end{align}
where $\mathcal{F}(\iota)  = (\cos \iota+1) (\cos \iota+5)^2/72$. Recently,
\citet{addison_gracia-linares_2019} used the same criterion to identify
surviving BBHs around SMBH. However, the criterion is derived using Newtonian
forces.  Its efficacy in the relativistic case is not yet testified.

\begin{figure}[ht!]
\includegraphics[width=0.5\textwidth]{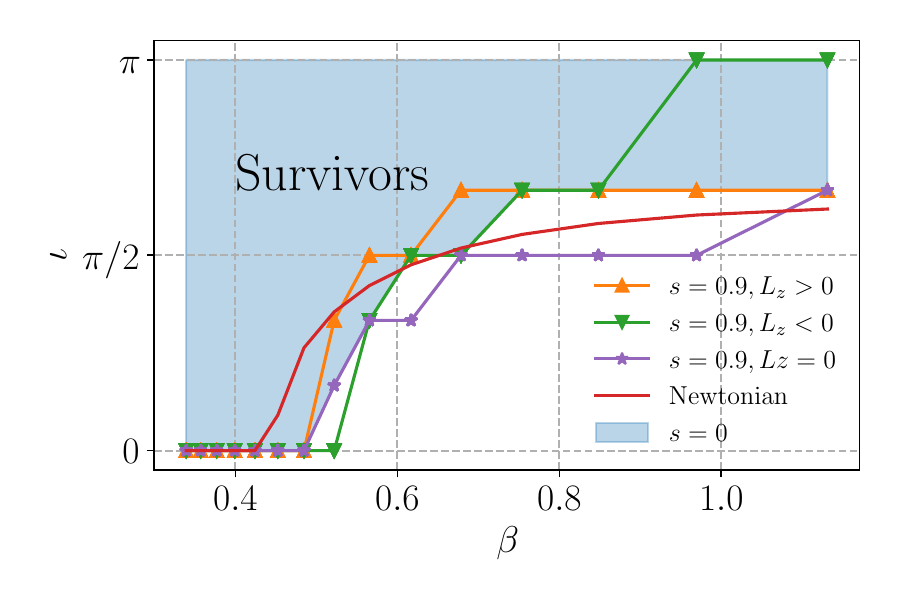}
\caption{
Comparing the Newtonian criterion (red solid line) for the survival of BBHs with the results
of our relativistic simulations. The upper and down triangles refer to the simulations with an
outer orbit inside the equatorial plane of a Kerr SMBH (with a spin parameter of $s=0.9$). 
The difference is that the upper triangles have $L_z>0$,
while the lower triangles have $L_z<0$. The purple stars show the results of inclined orbits,
which have $L_z=0$. The blue shaded region corresponds to the surviving BBHs around a Schwarzschild SMBH.
\label{fig:InclD}}
\end{figure}

Figure~\ref{fig:InclD} compares the Newtonian criterion (red solid line) with
the results derived in our relativistic simulations (lines with symbols as well
as the blue shaded region). In the plot, the survivors are lying either above a
demarcation  line or in the shaded area, depending on their initial conditions.
Interestingly, we find good agreement in most of the cases. Significant
deviation appears when the penetration factor $\beta$ exceeds $0.9$, 
in the simulations of $s=0.9$ and $L_z<0$ (green line with
down triangles).  In these
cases, we see that the BBH in the relativistic simulation is more stable than
that in the Newtonian one. The stability in the relativistic simulation could
be attributed to a faster rotation of the tidal field with respect to the rest
frame of the BBH.

\subsection{Eccentricities of the surviving BBHs}\label{sec:eccentricity}

\begin{figure}[ht!]
\includegraphics[width=0.5\textwidth]{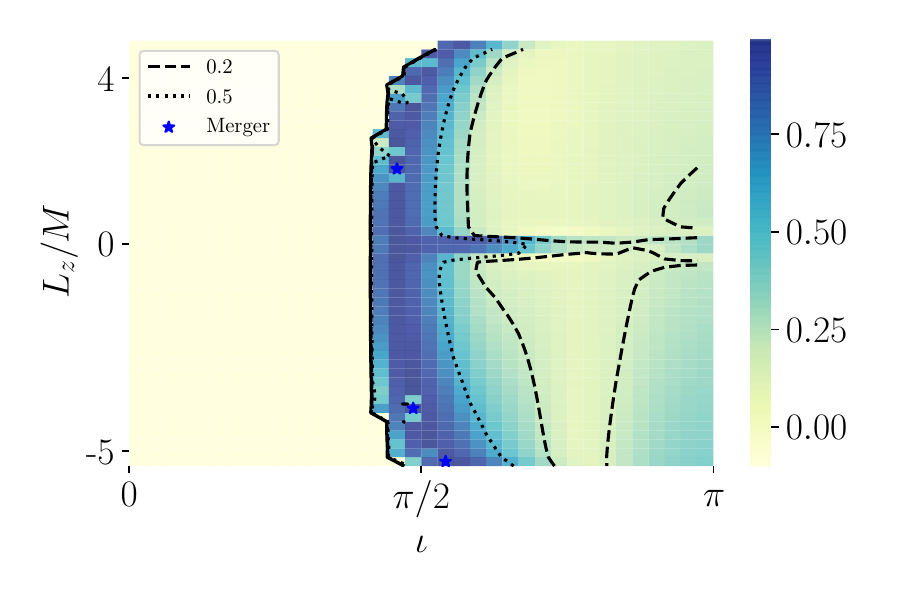}
\caption{Final eccentricity of the inner binary as a function of the
initial values of $L_z$ and $\iota$.
The solid line marks the boundary between the surviving and the disrupted binaries. 
The dashed and the dotted lines show, respectively, the contours for $e_{\rm fin}=0.2$ and $e_{\rm fin}=0.5$.  
The other parameters are the same as in Figure~\ref{fig:compF12}.
\label{fig:efin}}
\end{figure}

Beside semi-axis, the eccentricity of an inner BBH also varies during the
interaction.  Figure~\ref{fig:efin} shows the final eccentricity
after the BBH has completed the interaction and traveled to a distance of $r=100M$ from the SMBH.   

First, we find that the final eccentricity $e_{\rm fin}$ can exceed $0.5$ when
the inner BBH is inclined with respect to the rotational axis of the LIF, i.e.,
when $\iota\simeq\pi/2$. The cause is similar to the classical Von
Zeipel-Lidov-Kozai mechanism, in which the inner binary trades its inclination
for eccentricity \citep{vonZeipel1910,Lidov1962,Kozai1962}. The  major
difference, however, is that the tidal tensor in our LIF is asymmetric and
contains non-diagonal terms, while in the classical scenario the tidal tensor
is diagonal and is symmetric in the two azimuthal directions.  Second, close to
$L_z=0$, some BBHs with $\iota\ne\pi/2$ are also excited to high
eccentricities.  The cause, as has been shown in Figure~\ref{fig:compF12}, is
the appearance of large off-diagonal terms in the tidal tensor as the binary
reaches the polar regions of the Kerr SMBH. Third, we find that slightly more BBHs at $L_z<0$ than
those at $L_z>0$ are excited to an eccentricity of $e_{\rm fin}>0.2$. The
reason is that the rotation velocity $\omega$ of the LIF is slightly smaller
when $L_z<0$.  Therefore, in the rest frame of the BBH, the tidal field rotates
slower,  since this relative rotational velocity is $\omega+\omega_b$ for
$\iota>\pi/2$.

\begin{figure}[ht!]
\includegraphics[width=0.5\textwidth]{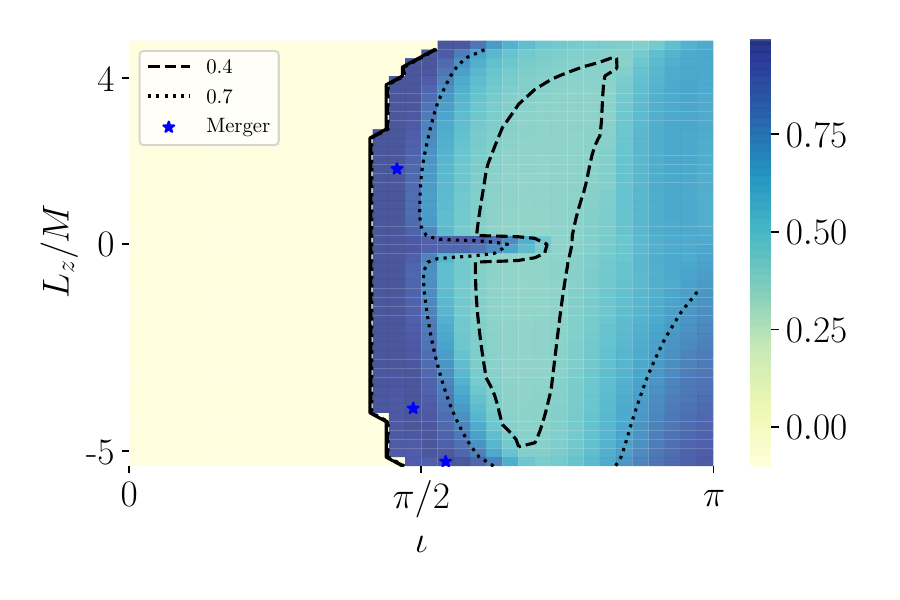}
\caption{Same as Figure~\ref{fig:afin} but showing the maximum eccentricity
	of the inner BBH during the encounter with the Kerr SMBH.
\label{fig:emax}}
\end{figure}

We notice that during the close encounter with the central SMBH, a BBH can
temporarily reach a very high eccentricity. Figure~\ref{fig:emax} shows
the maximum eccentricity recorded during the interaction of each BBH with the SMBH. 
We find that most surviving BBHs have reached an eccentricity as high as $e_{\rm max}=0.4$,
and many have exceeded $e_{\rm max}=0.7$. Such large eccentricities are expected to
result in GW bursts, which will be further discussed in Section~~\ref{sec:GWb}.

Significant variation of the eccentricity of the inner binary is also found
in our earlier study of a BBH moving along a circular orbit in the equatorial
plane \citep{paper2022}. However, it happens only when the binary is much
closer to the SMBH, e.g., $r<3M$. In the current work, the closest distance
between the BBH and the SMBH is $10M$. Nevertheless, significant variation of
eccentricity is seen in a large fraction of the parameter space (see
Figure~\ref{fig:emax}). Such an easier excitation of eccentricity highlights
the dynamical consequence of the asymmetric tidal tensors at the places outside
the equatorial plane.  It also indicates that BBHs reaching a distance of
$r\sim10M$ from a SMBH are more susceptible to merger than previously thought.

\subsection{Lifetimes of the surviving BBHs}\label{sec:GW}

We have seen that the eccentricities of the surviving BBHs are excited during
their close encounters with the Kerr SMBH.  The lifetime of a compact
binary is sensitive to the eccentricity since the GW radiation timescale
is
\begin{equation}
\begin{aligned}
    T_{\mathrm{gw}}&= \frac{5a^4 F(e) }{256 m_1 m_2 m_{12} }
	\simeq 5.2\times 10^4 \left(\frac{a}{a_0}\right)^4 F(e)~{\rm yr},\label{eq:merger}
\end{aligned}
\end{equation}
where $F(e) = (1 - e_0^2)^{7/2} (1 + 73 / 24  e_0^2 + 37 / 96 e_0^4)^{-1}$
\citep{peters_1964} and $a_0=20,000m_{12}$ is the initial semimajor axis. 
As a result, we expect a shorter lifetime for the surviving binaries.

\begin{figure}[ht!]
\includegraphics[width=0.5\textwidth]{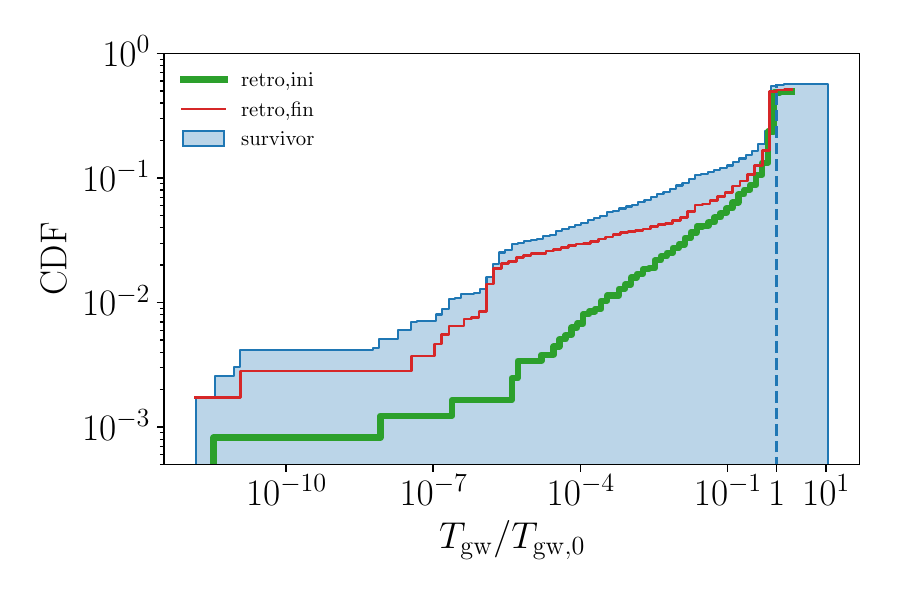}
\caption{Cumulative distribution function of the ratios $T_{\rm gw}/T_{{\rm gw},0}$ for the
BBHs surviving the interaction with the SMBH (blue histogram), where
$T_{{\rm gw},0}\simeq2\times10^5~{\rm yrs}$ is the GW-radiation timescale of the original BBH
and $T_{\rm gw}$ is the corresponding timescale after the interaction. 
The blue dashed vertical line marks the boundary of $T_{\rm gw}/T_{{\rm gw},0}=1$. The
green thick (red thin) histogram corresponds to the BBHs which are
initially (finally) counterrotating relative to the LIF.
\label{fig:Tgw}
}
\end{figure}

The lifetimes of the post-interaction BBHs are shown in Figure~\ref{fig:Tgw},
where we have assumed a random distribution for the initial orientation of the
BBHs in the FFF.  We find that in $54\%$ of the cases, the interaction results
in a BBH with a shorter lifetime.  Only in about $3\%$ of our simulations do we
find a longer lifetime for the BBH after the interaction.  Noticeably, in about
$12\%$ ($4\%$) of the case, the lifetime is shortened by a factor of $10$
($10^4$).

The green thick (red thin) histogram in Figure~\ref{fig:Tgw} shows the BBHs
which are counterrotating ($\iota>\pi/2$) with respect to the LIF before
(after) the interaction with the SMBH. At $T_{\rm gw}/T_{{\rm gw},0}\la0.1$, we
see a significant increase of the number of retrograde BBHs after the
interaction has completed. This result indicates that many BBHs with
significantly shortened lifetimes are originally corotating ($\iota<\pi/2$)
with the LIF, but their orbital orientations flip during the interaction with
the SMBH. Such an orbital flipping will affect the effective spin of the
binaries \citep[also noticed in][]{Joseph2019MNRAS}, which will be further discussed in section~\ref{sec:effspin}.

In particular, in about $f_m\simeq0.17\%$ of our scattering experiments, we find
that the BBHs coalesce during the encounter with the SMBH. In another
$f_i\simeq24.1\%$ of the cases, the lifetimes of the BBHs are shortened to a
value significantly less than the typical dynamical timescale of the galactic
nucleus, $r_i/\sigma\simeq4\times10^4$ yrs, where
$\sigma\simeq75~\mathrm{kms^{-1}}$ is the stellar velocity dispersion of the
nucleus of a Milky-Way-like galaxy \citep{TremaineEtAl02} and $r_i=GM/\sigma^2$
is the radius of gravitational influence of the SMBH.  The latter BBHs, because
of their shortened lifetimes, will also coalesce before they come back to
interact with the SMBH again.  Taking the above two types of binaries into
account, we derive a lower limit for the merger rate of BBHs, 
\begin{equation}
\begin{aligned}
\mathcal{R} &= p\Gamma_{\mathrm{BHB}}\cdot n_{g}\cdot f \\
&\simeq(0.78-1.94)\times10^{-2} \mathrm{Gpc^{-3}yr^{-1}},
\end{aligned}
\end{equation}
where $p\Gamma_{\mathrm{BHB}}\sim
(1.6-4)\times10^{-9}~\mathrm{yr}^{-1}~\mathrm{galaxy}^{-1}$ is the supply rate
of compact BBHs ($a\sim a_0$) to the SMBH which was estimated in
\citet{han_chen_2018}, $n_g\sim 2\times10^{7}~\mathrm{Gpc}^{-3}$ is the number
density of galaxies \citep{Conselice2005apj}, and $ f=(f_i+f_m)\simeq 0.24$.
The real merger rate should be higher since many surviving BBHs could come back
and interact with the SMBH again. A more accurate estimation of the merger rate
requires a dynamical model which can self-consistently track the formation and
evolution of BBHs in nuclear star clusters \citep[e.g.][]{zhang23}. We will
incorporate such a model in our future work. 
 
\section{Discussions and conclusion}\label{sec:dis}

In this work, we simulated the close encounter ($r_p=10M$) of a Kerr SMBH with a BBH coming from a
parabolic outer orbit. We solved the equations of motion of the binary in its FFF, 
in which the dynamics is no longer highly relativistic. We showed that in the FFF, 
a BBH rotating in the same direction as the tidal field is more likely disrupted.
Those surviving binaries, mostly counterrotating with respect to the tidal field, are excited
by the tidal field to a large orbital eccentricity so that about $54\%$ of them end up with
a shorter lifetime. Our results have important implications for future GW observations,
which we now discuss.

\subsection{Multi-band GW bursts}\label{sec:GWb}

The reason why we consider $a_0\sim10^4m_{12}$ is that such a binary falls in the sensitive band of
LISA \citep[also see][]{2018CmPhy...1...53C,paper2022}. The frequency of the GWs is
\begin{equation}
\begin{aligned}
   f_{\mathrm{gw}}\sim 2f_{\mathrm{orb}} &=\frac{1}{\pi}\sqrt{\frac{m_{12}}{a^3}}\\
   &=0.23\mathrm{mHz} \left(\frac{m_{12}}{\mathrm{25}M_{\odot}}\right)^{-1}
   \left(\frac{a}{a_0}\right)^{-3/2},\label{eq:fgw}
   \end{aligned}
\end{equation}
where we have assumed $e=0$. The amplitude of the GW radiation is
\begin{equation}
\begin{aligned}
    h_{\mathrm{gw}}&\sim \sqrt{\frac{32}{5}} \frac{m_1m_2}{d\,a}\\
    &\simeq 3.6\times 10^{-21}\left(\frac{d}{\mathrm{10kpc}}\right)^{-1}
    \left(\frac{a}{a_0}\right)^{-1}
    \label{eq:hgw}       
\end{aligned}
\end{equation}
\citep{peters_mathews_1963,xuan2023}, where $d$ is the luminosity distance of
the source. The latter equation indicates that under normal circumstances it is
difficult for LISA to catch such a BBH unless the binary is inside the Milky Way.

However, the increase of the inner orbital eccentricity during the encounter
with the SMBH, as is shown in Figure~\ref{fig:emax}, could shift the main power of the GW radiation into the
LIGO/Virgo band.  It is known that the frequency where the GW spectrum peaks
correlates with the characteristic frequency of the orbital pericenter,
$\sqrt{m_{12}/r_{p,b}^3}/\pi$, where $r_{p,b}$ is the pericenter distance of the inner orbit
\citep{wen_2003}. In our simulations, we found that $r_{p,b}$ could be as small
as $0.002a_{0}$ in about $0.3\%$ of the cases. Therefore, the GW frequency
could increase to
\begin{equation}
	f_\mathrm{{gw},p}\simeq 2.52\mathrm{Hz} \left(\frac{m_{12}}{\mathrm{25}M_{\odot}}\right)^{-1} \left(\frac{r_{p,b}}{0.002a_0}\right)^{-3/2},
\end{equation}
a frequency to which LIGO/Virgo are sensitive. Moreover,
the GW amplitude of an eccentric BBH can be estimated with 
\begin{equation}
	h_{\mathrm{gw},p}\simeq 1.8\times 10^{-22}\left(\frac{d}{\mathrm{100Mpc}}\right)^{-1}\left(\frac{r_{p,b}}{0.002 a_0}\right)^{-1}
\end{equation}
\citep{xuan2023}. Such an amplitude becomes accessible by the current LIGO/Virgo detectors.

Therefore, even though we started with a LISA BBH, the binary could exit the
LISA band and enter the LIGO/Virgo band during its closer encounter with the
central SMBH. Note that the encounter is short, lasting only a couple of
orbital periods of the BBH, since $\omega_b/\omega$ is slightly greater than
unity.  After the encounter, the eccentricity of the BBH returns to a relatively
low value as is shown in Figures~\ref{fig:efin}, which indicates that the
binary returns to the LISA band again.  Such an excursion would have produced
a couple of detectable bursts in the LIGO/Virgo band.

We have seen that when the inner binary enters the LIGO/Virgo band, 
the c.m. of the binary also reaches the pericenter of the 
outer orbit. This passage also produces a burst of GW radiation, which we refer to as the 
``extreme-mass-ratio burst'' (EMRB).
Notice that the conventional picture of an EMRB involves only one
stellar-mass BH around an SMBH \citep{Rubbo2006ApJEMRB,HopmanFreitagLarson07,YunesEtAl2008,BerryGair2013}, but in our case we have two stellar-mass BHs \citep{Joseph2019MNRAS}.
Following the earlier calculation of the GW signal of an eccentric binary, we find that 
the EMRB has a characteristic frequency of
\begin{equation}
	f_\mathrm{{EMRB}}\simeq 0.13\mathrm{mHz} \left(\frac{M}{4\times10^6M_{\odot}}\right)^{-1} \left(\frac{r_{p}}{10M}\right)^{-3/2},
\end{equation}
and a typical amplitude of 
\begin{equation}
	h_{\mathrm{EMRB}}\simeq 3.1\times 10^{-21}\left(\frac{d}{\mathrm{100Mpc}}\right)^{-1}\left(\frac{r_{p}}{10M}\right)^{-1}.
\end{equation}
Such a signal is detectable by LISA. 

The above analysis of the GW signal indicates that our system may be a multi-band GW source.
When the inner BBH passes by the central Kerr SMBH, its inner orbital eccentricity could be excited
so that the binary produces high-frequency GW bursts which are detectable by LIGO/Virgo.
Meanwhile, its motion around the SMBH could also produce a low-frequency GW burst which falls 
in the LISA band. Our analysis suggests that LIGO/Virgo/LISA can detect such a source 
out to a luminosity distance of ${\cal O}(10^2)$ Mpc.

\subsection{Effective spin of merging BBH}\label{sec:effspin}

The effective spin of a BBH is a measurable quantity in GW signal
\citep[e.g.][]{ligo2016PhRvL}, and it is defined as
\begin{equation}
\begin{aligned}
	\chi_{\mathrm{eff}}:=&\frac{m_1 \mathbf{S_1}+m_2 \mathbf{S_2}}{m_{12}} \cdot \mathbf{ \hat L_{in}},\label{eq:effspin}
\end{aligned}
\end{equation}
where $\mathbf{S}_{1,2}$ are the dimensionless spin vectors of the two BHs, and
$\mathbf{ \hat L_{in}} $ is a unit vector aligned with the orbital angular
momentum of the BBH \citep{Damour2001,XeffPNNRPRD,XeffIMRPRL}.  It is suggested
that the distribution of $\chi_{\mathrm{eff}}$ could reveal the formation
channel of the LIGO/Virgo BBHs
\citep{Farr2017xeff,TalbotPhysRevDeff,Vitale2017CQGeff}.  We have seen in
Section~\ref{sec:GW} that the orbital angular momentum of a BBH could flip its
direction during the close interaction of the binary with the SMBH. According
to Equation~(\ref{eq:effspin}), the value of $\chi_{\mathrm{eff}}$ will change.

To understand the impact of the angular-momentum flip on observation, here we analyze the
distribution of $\chi_{\mathrm{eff}}$ during two phases when the GW is the strongest, i.e.,
(i) the burst phase when the eccentricity of a BBH is the highest and (ii) the final merger when
the BBH is safely in the LIGO/Virgo band.
To simplify the analysis, we assume that initially the spin axes are aligned with the
original direction of the angular momentum, $\mathbf{ \hat L_{in,0}}$,
so that $\chi_{\mathrm{eff},0}/S=1$. We also assume that during the interaction 
the direction and magnitude of the spin of each BH
is constant (in the FFF), since the term of spin-orbit coupling is only a factor of 
$(v/c)^{3/2}/0.1\sim{\cal O}(10^{-2})$ of the tidal force imposed on the BBH.
Therefore, the effective spin during the encounter with the SMBH can be calculated with
\begin{equation}
    \chi_{\mathrm{eff}}/S=  \mathbf{ \hat L_{in,0}}\cdot \mathbf{ \hat L_{in}},
\end{equation}
where $S=(m_1 |\mathbf{S}_{1}|+m_2 |\mathbf{S}_{1,2}|)/m_{12}$ is the
dimensionless spin of the BBH.  After the encounter ($r>100M$),
$\chi_{\mathrm{eff}}$ remains approximately constant
\citep{Racine2008PhysRevD}.

\begin{figure}[]
\includegraphics[width=0.5\textwidth]{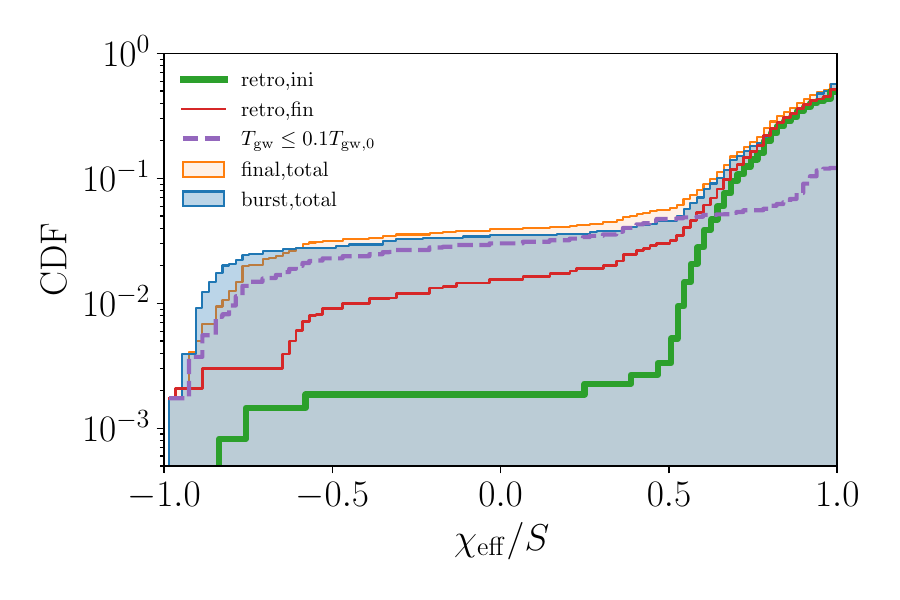}
\caption{Cumulative probability distribution of the effective spins of the inner binaries. 
The yellow and blue histograms show the effective spins measured at, respectively, the final merger 
and the burst phase. The green and red histograms have the same meanings as in Fig.~\ref{fig:Tgw} but for the effective spins
of the merging BBHs. Additionally, the purple dashed histogram shows the distribution of the BBHs whose lifetimes 
are shortened by more than a factor of $10$ due to their interaction with the SMBH.
\label{fig:effF}}
\end{figure}

The cumulative distribution of $\chi_{\mathrm{eff}}/S$ is shown in 
Figure~\ref{fig:effF}. We find that during the burst phase and the final merger,
the effect spin covers a wide range, unlike its initial distribution only at $\chi_{\mathrm{eff}}/S=1$.
About $4\%$ of the surviving BBHs have $\chi_{\rm eff}\leq -0.5$ when they merge. This
fraction is about four times higher than the percentage found previously
in Newtonian simulations \citep[see Fig.~(7) in][]{Joseph2019MNRAS}.
The significant difference highlights the necessity of carrying out 
relativistic three-body simulations like ours.

Figure~\ref{fig:effF} also shows that at $\chi_{\rm eff}\leq 0.5$, a relatively small fraction
of BBHs initially have retrograde inner orbits. 
This result corroborates our earlier observations that retrograde inner binaries 
in general are more stable than those prograde ones.

\subsection{Caveats}\label{sec:cav}

So far, we have focused on the systems with parabolic outer orbits.  However,
our method also applies to the BBHs on eccentric or hyperbolic outer orbits.
In particular, BBHs moving on eccentric outer orbits could be produced by a
dynamical process called ``tidal capture''
\citep{2018CmPhy...1...53C,addison_gracia-linares_2019}.  Such a captured BBH
would interact with the central SMBH multiple times before it is either tidally
disrupted or driven to merger. To study the final outcome, we need to further
consider the cumulative loss of the energy and angular momentum of the outer
orbit during successive close encounters. 
 
When choosing the initial conditions, we only considered a variation of the
orientations of the inner and outer orbits while keeping the other parameters
fixed.  It has been reported by previous works that the initial values of the
ascending node and the phase of the inner binary also play a important role in
determining the outcome \citep{addison_gracia-linares_2019,Joseph2019MNRAS}.
Dedicated simulations are needed to cover such a parameter space of higher
dimension.

When estimating the frequency and amplitude of the GW burst emitted during a
close encounter of two BHs, we made a simplification that the observable
quantities are similar to those of a circular binary whose orbital semimajor
axis is the same as the closest distance during the encounter.  However, a
burst signal contains multiple GW frequencies and, in fact, we have considered
in this work only the most powerful GW mode. Taking the full GW spectrum into
account could enhance the signal-to-noise ratio and help identify burst events
in future GW observations.

Although the PN formalism adopted in this work can appropriately treat the
evolution of the BH spin in the FFF, we find that in most cases the
corresponding term (1.5 PN) remains small relative to the GE force. This is the
reason why we only considered the variation of the orbital angular momentum
when we calculated the effective spin of the inner binary. Only in a couple
of rare cases, where the BBHs coalesce during their close encounter with the
SMBH, did we find that the 1.5PN term is no longer negligible. In these cases,
the BHs become so close to each other that higher order PN corrections, such as
those up to 3.5 PN order \citep{2017PhRvD..96b3017W}, need to be included in
the calculation.

Despite these caveats, our series of works \citep[also see][]{paper2022} have
established a framework which can simulate the dynamical evolution of a compact
binary moving on an arbitrary orbit around a Kerr SMBH. Further improvements,
as described above,
will enable us to derive long-term evolution of the triple system and construct
accurate waveforms, which are very much needed for the detection such sources
in GW observation.

\section{Acknowledgement}
This work is supported by the National Key Research and Development Program of
China Grant No. 2021YFC2203002 and the National Natural Science Foundation of
China (NSFC) grant No. 11991053.
\bibliography{ref,bibbase,biblio}{}

\bibliographystyle{aasjournal}

\end{document}